# Fault Signature Identification for BLDC motor Drive System -A Statistical Signal Fusion Approach

Tribeni Prasad Banerjee[1], Susanta Roy[2], B. K. Panigrahi[3]

[1]Dept. of Electronics and Communication Engineering, Dr. B.C. Roy Engineering Collage, Durgapur-6, India
[2]Electrical Engineering Department, Jadavpur University, Kolkata 700 108, India
[3]The Centre for Automotive Research and Tribology (CART)
tribeniprasad.banerjee@bcrec.ac.in, raysusant@gmail.com, bkpanigrahi@ee.iitd.ac.in

*Abstract* — **A hybrid approach based on multirate signal processing and sensory data fusion is proposed for the condition monitoring and identification of fault signal signatures used in the Flight ECS (Engine Control System) unit. Though motor current signature analysis (MCSA) is widely used for fault detection now-a-days, the proposed hybrid method qualifies as one of the most powerful online/offline techniques for diagnosing the process faults. Existing approaches have some drawbacks that can degrade the performance and accuracy of a process-diagnosis system. In particular, it is very difficult to detect random stochastic noise due to the nonlinear behavior of valve controller. Using only Short Time Fourier Transform (STFT), frequency leakage and the small amplitude of the current components related to the fault can be observed, but the fault due to the controller behavior cannot be observed. Therefore, a framework of advanced multirate signal and data-processing aided with sensor fusion algorithms is proposed in this article and satisfactory results are obtained. For implementing the system, a DSP-based BLDC motor controller with three-phase inverter module (TMS 320F2812) is used and the performance of the proposed method is validated on real time data.**

## I. INTRODUCTION

Fault detection qualifies as a critical area of research in safety-critical systems such as aircraft, hybrid vehicle, industrial robotics application, nuclear process plant, and chemical reactor. Numerous methods have been developed to detect faults in electric machines such as induction motors (see for example [1 - 5]). The short circuit faults [6] using a fault tolerant controller has been investigated. In [7], detection of mechanical faults including rotor eccentricity and defects in permanent magnets of an inverter-fed surface mounted PMSM (Permanent Magnet Synchronous Motor) has been undertaken by using current signature analysis of line current. Now various signal processing and pattern recognition techniques have been employed to extract useful features of fault currents and the present state-of-the-art amounts to classify faulty signals of the motor drives. The signal processing methods consist of Fast Fourier Transform (FFT) [8], Short Term Fourier Transforms (STFT) [9], and Wavelet Packet Transforms [10]. However, non-periodic, wideband and non-stationary signals can be generated in faulty motor drives. In addition, the bandwidths of these fault-generated signals may lie well outside the perceptibility of the current protection techniques. Signal transformation based diagnostic techniques for drives cannot be in the acceptable region [10]. In fault diagnosis various expert strategies are the largest application domain [11]. The pattern recognition techniques [12], particle swarm optimization approach [13], are also used for faults diagnosis scheme. Even new statistical approaches like Support Vector Machines (SVMs) [14, 15]. Functional SVMs [16] has been used to enhance the accuracy of fault feature classification. There are different types of diagnosis methods like hardware-redundancy methods, knowledge based and signal processing methods or model-based approaches [17-19] are well established. Now there are various recent expert techniques such as Swarm Intelligence [20], Neural Networks [21], and hybrid neuro-fuzzy approaches [22] [23] which are more popular for their nonlinearity and better approximation capability but still we observe that most of the fault diagnostic techniques have been developed and implemented for induction motors. This makes it a key challenge to make some contribution in the BLDC motors' online current signature analysis based on our intelligent hybrid technology. In most of the previous works, the steady-state behavior of faulty drives has been focused. These works did not only concentrate on the theoretical but mostly on the practical implementation of the hybrid fault aware control system by signal processing technique based sensor fusion and signature identification hybrid approach shows in realistic approaches in the simple implementation point of view as well as it reduces the complexity of the system.

The STFT has its own drawback because of its window size because of that we only analysis the signal with some degree of precision. Thus, wavelet transform was been introduced to overcome the fixed length problem. Whereas, the performances of STFT and wavelet based hybrid techniques for the signal frequency monitoring show enhancement of the results [24]. In our proposed model a hybridization of intelligent predictive controller with STFT and multirate signal processing is used, in order to increase the efficiency of signal processing operations. After A/D conversion of the machine current and voltage signal, the intelligent processing unit transforms the signal through STFT techniques combining Blind source separation (BSS) and detects the change in source signals. The problem of change detection in multi sensor data and frequency information provided by different sensors are important issues for future signal signature analysis. In order to collect the required data for designing and testing the proposed diagnostic technique, an experimental setup for data acquisition and signal pre and post processing has been developed to mimic the electrical faults with the help of ***TMS 320F2812*** DSP processor. It allows for faults to be introduced in a controlled manner. The proposed hybrid diagnostic technique has been successfully integrated in a

BLDC motor with the DSP Processor as illustrated in Figure 1. The control signal and pre and post processing technique has been applied to the test the performance of the system with real time environment like electromechanical valve used in ECS unit for controlling the ambient of civil aircraft as shown in Figure 7. The performance of the novel hybrid technique has been evaluated through simulation and experimental results.

Rest of the paper is organized in the following way. Section II is mostly describing the Blind Signal Processing method, the description of the proposed system and the various block configurations is described in Sections III. The experimental setup and the simulation results have been shown in Section IV and V. Finally the paper is concluded in Section VI.

## II. BLIND SIGNAL PROCESSING

Blind Signal Separation or Processing (BSS/BSP) deals with the problem of recovering multiple independent sources from their mixtures. BSS is very much similar to the Independent Component Analysis (ICA) [25]. ICA is one method, perhaps the most commonly used, for performing blind source separation techniques. The simple model for BSS assumes the existence of $M$ independent signals from various sensor and the $N$ number of noise signals mixed during the measurement period, where the noise is linear but of random type [26]. The problem that it is solving can be formulated statistically in the following way: given $M$-dimensional random variable vector $\mathbf{x}(t) = [x_1(t),...,x_M(t)]^T$ that arises from linear combination of the mutually independent components of $N$-dimensional unknown random variable $\mathbf{s}(t) = [s_1(t),...,s_N(t)]^T$ represented mathematically as:

$$\mathbf{x}(t) = \mathbf{A}\mathbf{s}(t) \quad t = 1, 2, ..., M, \quad (1)$$

where $\mathbf{x} \in \mathbf{R}^M$, $\mathbf{s} \in \mathbf{R}^N$ and $\mathbf{A}$ is an $M \times N$ mixing matrix. Here $\mathbf{R}$ denotes the field of real numbers. The class of algorithms that can handle such a problem also falls under the category of ICA. When the number of the mixtures is equal to that of the sources (i.e. $M = N$), the objective can be refined to find an $N \times N$ invertible square matrix $\mathbf{W}$ such that:

$$\mathbf{u}(t) = \mathbf{W}.\mathbf{x}(t), \quad t = 1, 2, ..., N, \quad (2)$$

where the components of the estimated source $\mathbf{u}(t) = [u_1(t),...,u_N(t)]^T$ are mutually independent as much as possible. This must be done as accurately as possible with the assumption that no more than one source has a Gaussian distribution. Current algorithms can meet this objective within a permutation and scaling of the original sources. In ICA we attempt to optimize the following objective function with respect to the unmixing matrix $\mathbf{W}$ such that

$$L(u, \mathbf{W}) = \sum E[\log p_{u_i}(u_i)] - \log |\det(\mathbf{W})|, \quad (3)$$

where $E[.]$ represents the expectation operator and $p_{u_i}(u_i)$ is the model for the marginal Probability Density Function (PDF) of $u_i$, for all $i = 1, 2, ..., n$. Normally, matrix $\mathbf{W}$ is regarded as the parameter of interest and the PDFs of the sources are considered to be nuisance parameters. In effect, when correctly hypothesizing upon the distribution of the sources, the Maximum Likelihood (ML) principle leads to estimating functions, which in fact are the score functions of the sources [27]. Blind source separation consists in the recovering of the various independent sources exciting a system given only the measurements of the outputs of that system. BSS has become a mature field of research with many technological applications in areas such as wireless communications, antenna processing, and speech processing. The recent successes of BSS may be also used in mechatronics signal processing. The issue of inferring the nature of unknown endogenous sources from exogenous measurements has constantly been a major apprehension in this field. The works in this domain have already proved that BSS provides new solutions for vibration and noise analysis. However, when it comes to deal with mechanical signals, which are typically characterized by an excessive complexity, BSS faces a number of difficulties which seriously hinder its feasibility [28, 29].

## III. PROPOSED HYBRID MODEL FOR THE CONTROLLER IN AN INDUSTRIAL ENVIRONMENT

The change detection of a complex electromechanical system context of fault diagnosis by signal signature is a typical signal segmentation problem. Such type of detection system helps the system to prevent the subsequent occurrence of more catastrophic failure. Such a signal segmentation tool helps to increase the reliability and of the model by reducing the number of auto shutdowns that are usually unnecessary. In order to achieve high performance, we design our controller more adaptive with response to various load and external fault related signal. Wavelet Packet Transform (WPT) based neuro-fuzzy algorithm [10] has been developed and successfully implemented for stator electrical faults of a line-fed IPM motor and condition based monitoring has been done in very recently. In this paper, we developed a signal signature based separation based diagnostic methods for a BLDC motor. For comparing the proposed diagnostic technique results, an experiment has been developed to mimic the electrical faults and the data has been collected from the system. Our proposed methods are novel because the system has been integrated with high speed DSP processor that process the signal and classify the signal into various fault signal signatures. The proposed technique is tested and on a laboratory prototype motors using the digital signal processor (DSP) board of Texas Instrument *(TMS 320F2812)*. The performance of this novel hybrid technique has been evaluated through simulation and experimental results. The motion tracking and control of BLDC motor in a nonlinear situation plays a very crucial role in aircraft control system. In this paper we are mainly concentrated on the signal signature based automatic preprocessing in dynamic situation. The model has been designed in MATLAB Simulink, and testing on the on-board DSP system. The architecture of the proposed hybrid intelligent controller has been shown schematically in the Figure 1.

Now measurement techniques in combination with advanced computerized signal acquisition and processing show new ways in the field of signal signature monitoring by the use of spectral analysis of operational parameters (e.g., FFT, STFT, WPT, etc.). Time-domain analysis using characteristic values to determine changes by trend setting, spectrum analysis to determine trends of frequencies, amplitude and phase relations, to detect periodical components of spectra are used as evaluation tools.

In many situations, signal monitoring methods were utilized for incipient fault detection [30]. However, we are more concerned about the signal separation scheme that has been proposed in this paper. This model has the following processing sections as shown in Figure 1 schematically.

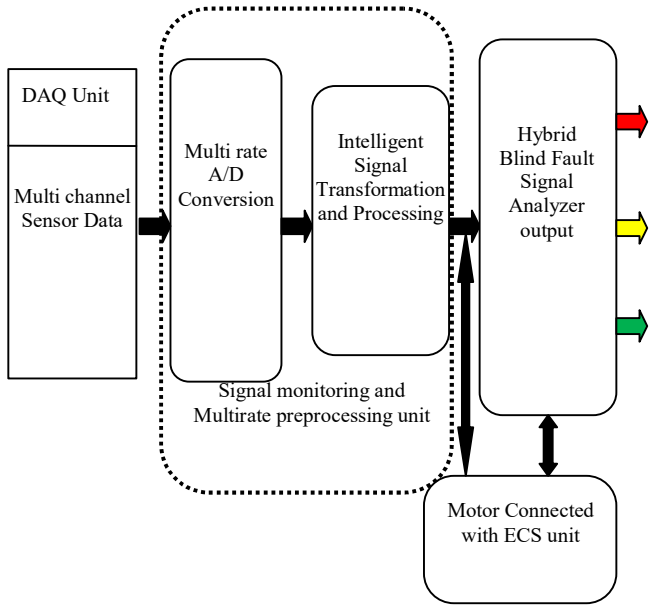

**Fig. 1.** The schematic architecture representation of the hybrid multi signal processing technique

**1. DSP Controller:** Thanks to recent digital signal processor (DSP) technology developments, motor fault diagnosis can now be done in real time fashion. This block basically takes the input current signal for controlling the motor speed with the dynamic changes of the sensory data and carefully determined based on practical issues such as noise content, phase unbalance, or frequency errors [23, 31] maintaining the complexity low enough for DSP-based real time implementation has been proposed [30]. Here we used the standard Texas Instrument TMS320F2810 kit for processing the data and taking the analog data from the system which significantly simplifies the detection algorithm. The user can generate periodic fault signatures with the help of the system.

**2. BLDC Motor:** The speed and the position of the BLDC motor has to control with the dynamic behavior of the sensor data. This kind motor is very attractive in servo and/or variable speed application since it can produce a torque characteristic similar to that of a permanent magnet DC motor while avoiding the problems of failure of brushes and mechanical commutation. But still electromechanical system coupled with motor and an alternative way of detecting faults in gears coupled to BLDC motors by monitoring either the motor current or voltage [32, 33].

**3. Automated Signal Classifier:** The major task of the blocks is to classify the signal into a specific class from that we can detect the system behavior in prior to the system is going to a faulty situation. Thus, the performances of the FFT and STFT-based diagnostic techniques shows also unsatisfactory results for BLDC motor as mentioned [10]. In this proposed model we implemented the hybrid multi rate blind signal processing techniques depending upon the sampling rate and the noise distribution [28]. The system generated data has been used to train the hybrid BSP algorithm. The online fusion of the pressure and temperature relation always to maintaining the system specified standard ambient. The online detection has a two part one is supervised with synthetic data. The schematic diagram of the internal structure has been shown in to Figure 1.From this simplified sensor signal we can easily processed for signature analysis and classification. The classified signal also fed to the system to taken care for comparison that the system is going to next state of damaging condition or better response. Then the fault controller output and the automated signal controller.

**4. Fault Predictor:** Due to the broad scope of the process fault diagnosis problem and the difficulties in its real time solution, various computer aided approaches have been developed over the years. The automation of process fault detection and diagnosis using expert system is an area of investigation concern with a real time system monitoring. However, this complete belief on human operators to cope with such abnormal events and emergencies has to several factors of a sophisticated complex real time embedded system. So the fault controller block has been introduces. This block is playing a crucial role into the controlling the DSP and the MOTOR also. The first role of the block is to test the system is going to fault or it has an incipient fault [30]. Second major responsibility of the block is data acquisition and predicts the fault signal for the control system.

**5. Predictive current controller:** Disturbances acting on the process, measurement noise and model-plant mismatch cause differences in the behavior of the process and of the model. In open-loop inverse-model control, this results in an error between the reference and the process output. In [34] explicitly provides an estimate of the probability of a fault for a set of known fault modes that is updated in real time, statistical confidence levels, the internal model control (IMC) scheme [35] is another way of compensating for this error. Figure 1 depicts the scheme. Here the control system has two inputs, the reference taken from signal classified and the fault controller output, and one output, the control action given to the DSP.

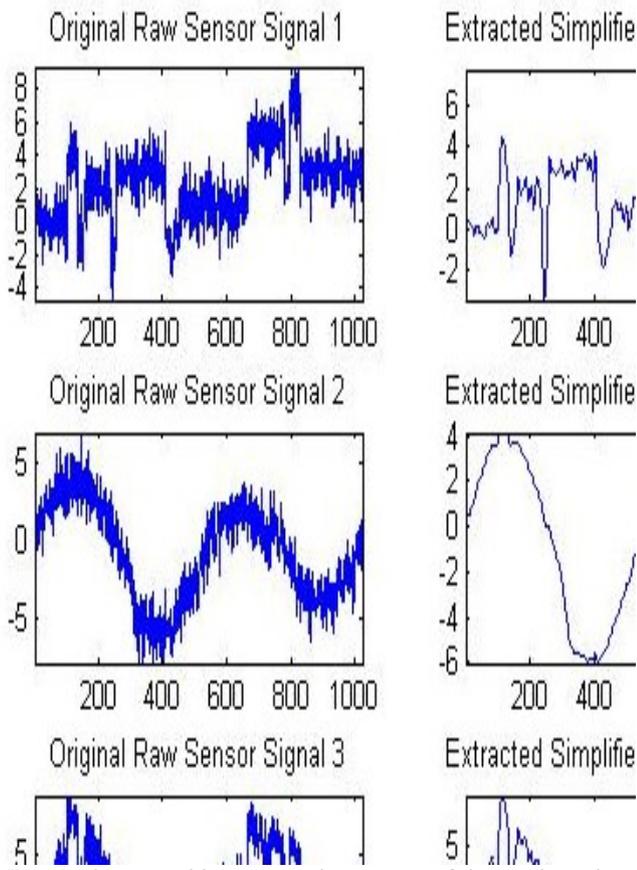

**Fig. 2.** The smoothing processing output of the random signal from the raw data collecting from data acquisition system

## III. SIMULATIONS AND RESULTS

In this paper for data acquisition purposes, a data translation board was used. Notably the whole data acquisition circuitry works only at a particular speed (in revolution per minute). Also, LabVIEW 8.2 version was used for data acquisition. Subsequently, the data acquired through this process is in the ".dat" format and is ready for analysis. Since the maximum frequency for the engine state can be 12 kHz, the nyquist criterion sampling frequency was employed at 50 kHz. Observably, a typical sample has a bit rate of 128 kb/s with multichannel. Figure 3 shows schematic diagram of the controller and the DSP processor setup.. In many proposed algorithms deal with the ambiguities of line current noise or sensor resolution errors and operating point−dependent threshold issues [36]. They have theoretically and experimentally validated faults into a motor. Hybrid system monitoring requires measurement or estimation of continuous state variables and tracing discrete states [30, 37].

In this paper, the present experimental setup for the entire process and simultaneously relates it with the actual parameters used in the analysis. For the present analysis, the data used is taken within a particular speed band which is 2800–3200 rpm. The simulation shows that the fusion based approaches and hybrid based approaches are more accurate signature oriented. And from the Figure 5 we can easily say that. Figure 5 shows that in various types of noise like random, uniform white, Gaussian are applied to the system and Figure 4 (a),(c),and 5(e) are the only fusion output which is mostly same but after the multirate processing and the fusion based approaches the signal signature output has been shown and in Figure 4 (b),(d),and 5(f).With the signal signature output the system can easily identify that there is some types of faults has been occurred because the prior knowledge of every signature has been known by the intelligent system. The system has been tested in various noise levels and sampling frequency and the results gives the appropriate signature as shown in the Figure 4 and Figure 5.

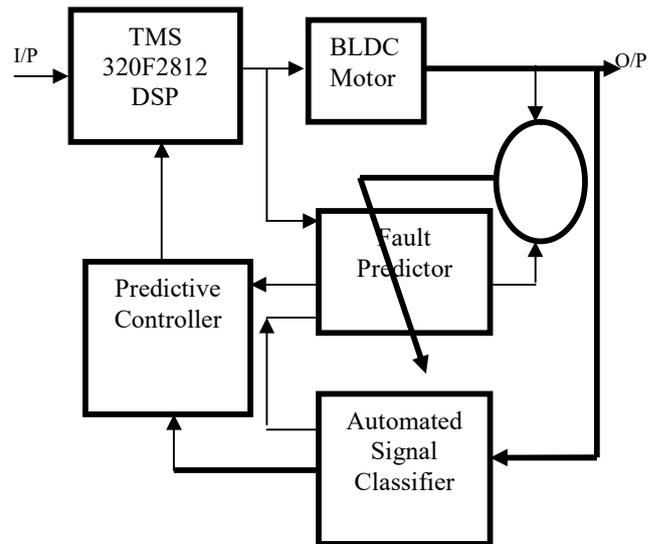

**Fig. 3.** Architecture of the proposed Intelligent Fault aware controller

The proposed technique did not require any harmonic contents analysis. The proposed technique is quite fast and easy to implement. It requires less computational memory for the online implementation. The dedicated examples specify the strength of the new approach for fusing and fault detection noisy sensor signals and multiple faulty sensors. This makes the new approach very attractive for solving such fusion problems and fault detection in real time operation. Especially for high performance operational system such as: rotorcraft, airplanes and process control systems. For the validation the test setup is for our proposed system operating frequency 50 Hz, and 28 volts supply at "0"phase shift, at first case sampling frequency ($f_s$) is 100.

A real time comparison has been made and it is established the better performance of the system as shown in the Figure 6. The experimental setup and the model of the hybrid controller have been shown in to the Figure 7.

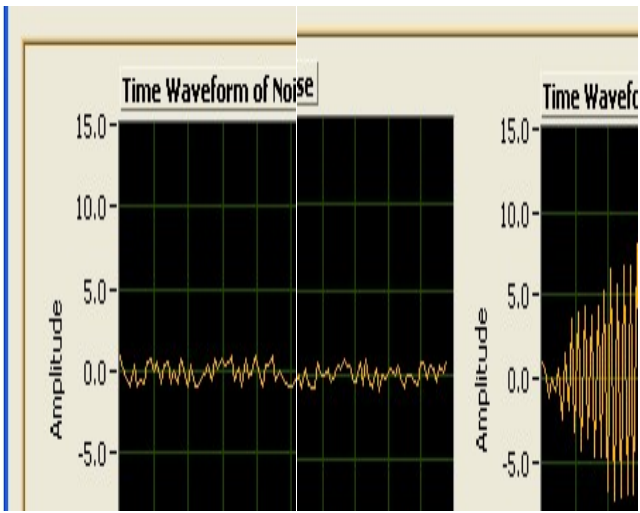

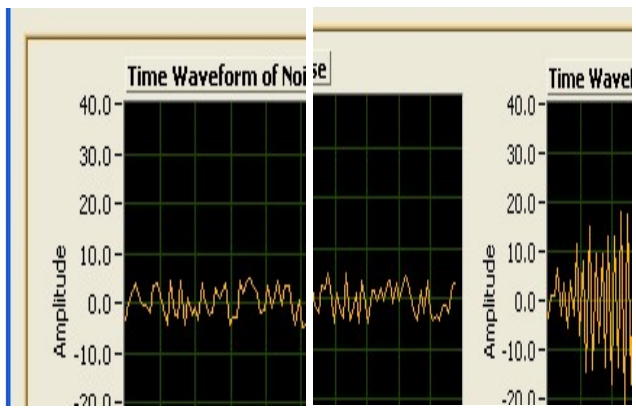

**Fig. 4.** Shows (a) sensor fusion input in random noise and (b) after processing the signature (c) at uniform white noise (d) the processing output signature of the signal at the various sampling rate simulation results of output

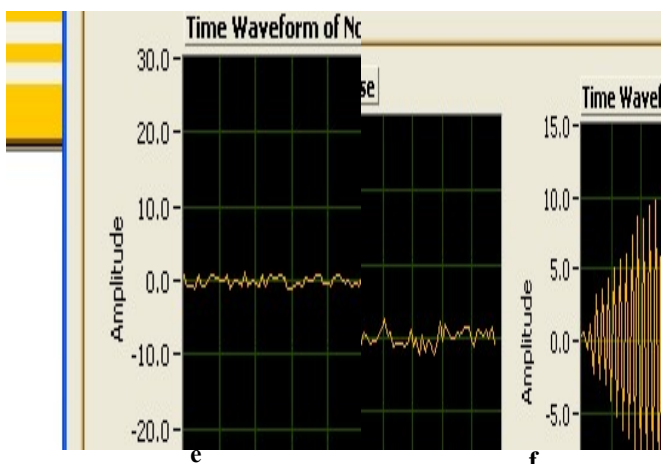

**Fig. 5.** Shows (e) at Gaussian noise (f) the processing output signature of the signal at the various sampling rate simulation results of output

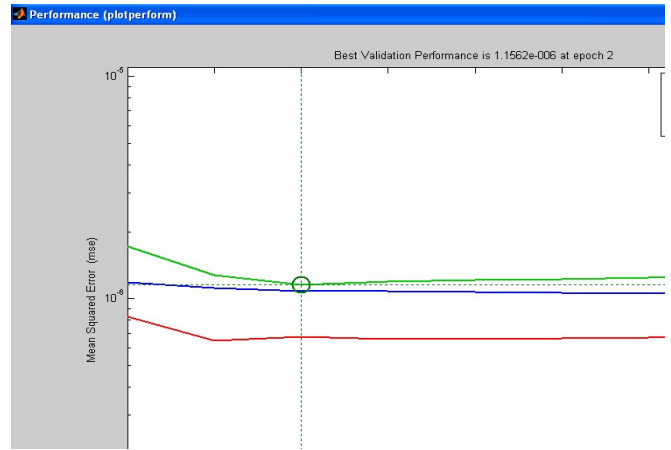

**Fig. 6.** Performance of the proposed hybrid system and the Fusion based approaches

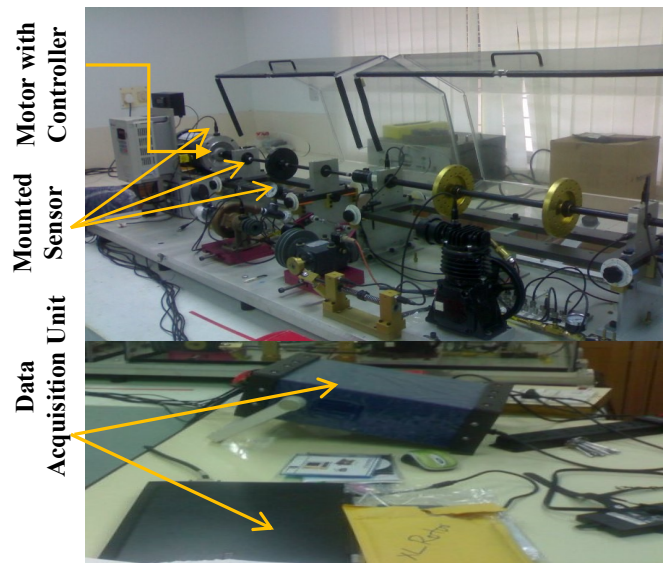

**Fig. 7.** Experimental setup of the intelligent fault signal simulator with data acquisition system

## IV. CONCLUSION

In this paper, a sensor fusion and multirate signal processing based hybrid technique is proposed and implemented in real time. Most literature dealing with fault diagnosis of ECS unit has applied their methods under laboratory environments and proposed that their algorithms can be used for online fault diagnosis under actual industrial environment. In the present research, the proposed technique carried out online fault diagnosis of ECS unit of an Aircraft under actual industrial environment with desirable accuracy and specified time limits. The proposed scheme has been tested in online fault diagnosis. Online testing was conducted with 3-GHz processor and 8-GB RAM memory. Analysis time of one unit is varied from 1.5 to 2 s with desirable accuracy and repeatability. Also, the proposed method works satisfactorily for engines having a maximum of two faults at the same time. The proposed technique is tested online on motors using the industrial fault monitoring setup. The test results show that the proposed technique is able to distinguish the signal signature before the signal feed to the controller so the controller can be more ambient intelligent to produce a false alarm because the purpose of an alarm is to alert the system operator to an ambient condition that requires immediate attention. An alarm is said to occur whenever the abnormal condition is detected and the alert is issued. An alarm is said to return to normal when the abnormal condition no longer exists. Developing an effective intelligent alarming system by this sensor fusion and signal processing techniques requires substantial commitments of effort, involving process engineers, control systems engineers, and production personnel. Methodologies such as expert systems can facilitate the implementation of an intelligent fault alarming system. Conventionally, this testing was done by an expert system in 60 to 120 s. An interesting topic for future work would be how the adaptive motion control can be implemented into an FPGA chip so the system size will be small as well as the performance of the dynamic behavior can be improved with high accuracy because it has a dedicated application into the aircraft.

ACKNOWLEDGMENT

The authors would like to thank CSIR for his research grant as senior research fellowship (CSIR-SRF Grant Ref no: 9/96(0690)2k11-EMR), in Jadavpur University to continue this research.